\documentclass[journal=jpccc,manuscript=article]{achemso}

\usepackage{chemformula} 
\usepackage[T1]{fontenc} 

\author{Binglin Zeng }
\affiliation
{School of Mechanical Engineering and Automation, Beihang University, 37 Xueyuan Rd, Haidian District, Beijing, China}
\alsoaffiliation
{Physics of Fluids Group, Department of Applied Physics and J. M. Burgers Centre for Fluid Dynamics, University of Twente, P.O. Box 217, 7500 AE Enschede, The Netherlands}

\author{Yuliang Wang }
\email{wangyuliang@buaa.edu.cn}
\affiliation{School of Mechanical Engineering and Automation, Beihang University, 37 Xueyuan Rd, Haidian District, Beijing, China}
\alsoaffiliation{Beijing Advanced Innovation Center for Biomedical Engineering, Beihang University, 37 Xueyuan Rd, Haidian District, Beijing, China}

\author{Xuehua Zhang}
\affiliation{Department of Chemical and Materials Engineering, University of Alberta, 12-211 Donadeo Innovation Centre for Engineering, Edmonton, Alberta, Canada}
\alsoaffiliation
{Physics of Fluids Group, Department of Applied Physics and J. M. Burgers Centre for Fluid Dynamics, University of Twente, P.O. Box 217, 7500 AE Enschede, The Netherlands}

\author{Detlef Lohse}
\email{d.lohse@utwente.nl}
\affiliation
{Physics of Fluids Group, Department of Applied Physics and J. M. Burgers Centre for Fluid Dynamics, University of Twente, P.O. Box 217, 7500 AE Enschede, The Netherlands}

\title[An \textsf{achemso} demo]
  {Solvent Exchange in a Hele-Shaw Cell: Universality of Surface Nanodroplet Nucleation}

\abbreviations{IR,NMR,UV}

\begin{document}

\begin{tocentry}
	\centering{\includegraphics{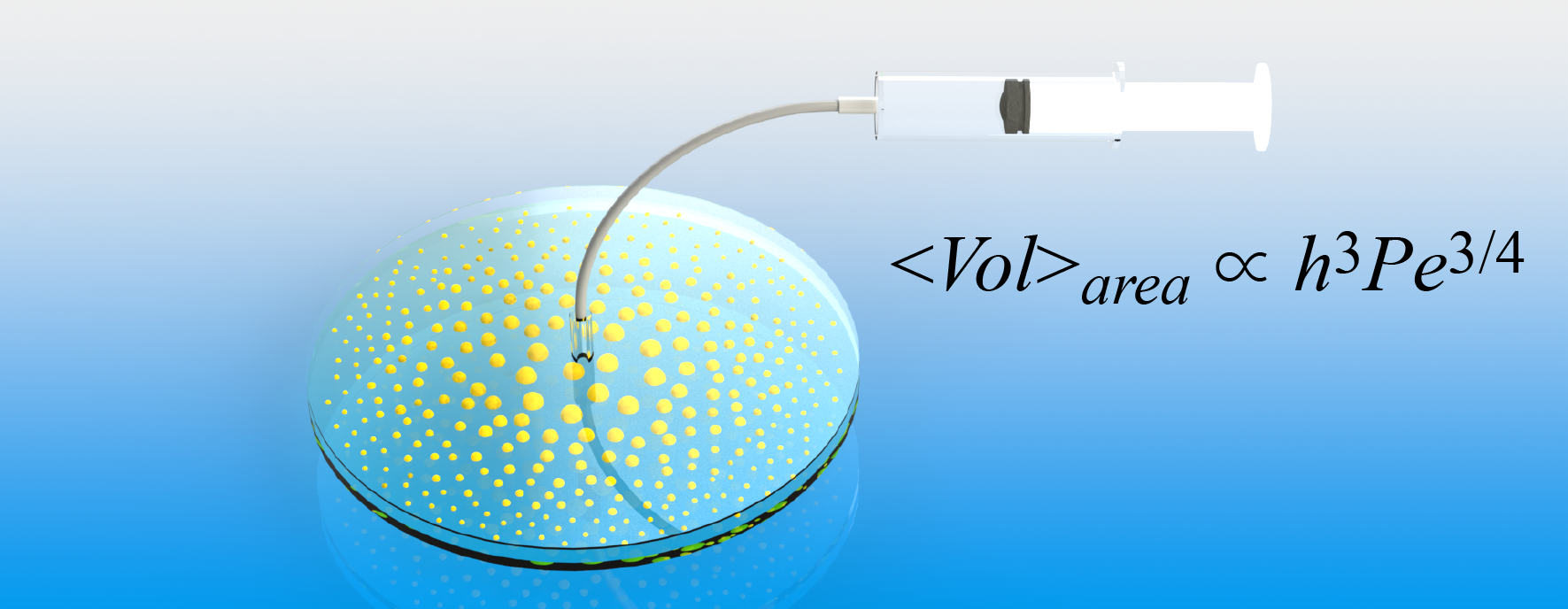}}
	\label{TOC Graphic}
\end{tocentry}
\begin{abstract}
Solvent exchange (also called solvent shifting or Ouzo effect) is a generally used bottom-up process to mass-produce nanoscale droplets.
In this process, a good solvent for some oil is displaced by a poor one, leading to oil nanodroplet nucleation and subsequent growth.
Here we perform this process on a hydrophobic substrate so that sessile droplets -- so-called surface nanodroplets -- develop, following the work of Zhang {\it et al.} [Zhang, X.; Lu, Z.; Tan, H.; Bao, L.; He, Y.; Sun, C.; Lohse, D. Proc. Natl. Acad. Sci. U. S. A. 2015, 122, 9253-9257]. In contrast to what was done in that paper, we chose a very well controlled Hele-Shaw geometry with negligible gravitational effects, injecting the poor solvent in the center of the Hele-Shaw cell, and characterize the emerging nanodroplets as function of radial distance and flow rates. We find that the mean droplet volume per area $\left.\langle Vol\right.\rangle _{area}$ strongly depends on the local Peclet number $Pe$ and follows an universal scaling law $\left.\langle Vol\right.\rangle_{area} \sim Pe^{3/4}$. Moreover, the probability distribution function of the droplet volume strongly depends on the local $Pe$ as well, regardless of the flow rates and radial distance, giving strong support to the theoretical model of the solvent exchange process developed in Zhang et al's work.
\label{abstract}
\end{abstract}

\section{Introduction}
The solvent shifting process -- also called solvent exchange -- is a simple and generic approach for mass-producing droplets or bubbles at solid--liquid interfaces. The droplet or bubble are only several tens to hundreds of nanometers in height, or a few femtoliters in volume \cite{lohse2015rmp,XiaoQianxiang2017langmuir,Xu2014Nanobubble,Xu2015AdvColloid,zhang2008droplets,zhang2018contactline}. In this process, a good solvent is replaced by a poor solvent, leading to nanodroplets or nanobubble nucleation and subsequent growth on the substrate. The approach has several potential applications, such that it can be used for liquid-liquid microextraction, diagnosis, drug production, extraordinary focusing, micromanufacture and among many others\cite{Strulson2012Nature, Chiu2009small, Shemesh2014pnas, Chen2015pccp, Meckenstock2014Oil, peng2014nanoemulsion, zhang2010nanodents, darwich2011,ma2014,Mart2015Sub,zhangran2019droplet,hung2018dynamic,gao2018selfremove}. When applied to oil dissolved in a good solvent, the solvent exchange process has the capability of mass-producing surface nanodroplets of oil on substrates in one step\cite{zhang2015pnas, zhang2014prl, peng2018,peng2016-coll, lu2016, lu2015}.

In ref.\ \cite{zhang2015pnas} the solvent exchange process was performed in a linear channel, with the flow rate $Q$ and the channel height $h$ as control parameters. The nanodroplet generation for seven different flow rates and three different channel heights between $h = 0.33$ mm and $h= 2.21$ mm was analysed. The main result was that the experimentally found mean droplet volume was consistent with the theoretical results derived in the same paper, namely that the area-averaged volume of the droplet $\left.\langle Vol\right.\rangle_{area}  \sim$ $h^{3}Pe^{3/4}$, where area-averaged volume $\left.\langle Vol\right.\rangle_{area}$ is defined by the total volume of droplets over a unit surface area, $Pe = Q/ (wD)$ is the Peclet number, defined by the flow rate, the width of the channel $w = A/h$ (where $A$ is the rectangular channel cross section), and the mass diffusivity $D$ of the oil, see figure 2G of that paper.

However, for the larger channel heights analyzed in that paper, major convective effects sets in, due to the density difference between the two solvents, leading to considerable inhomogeneities in droplet sizes. Indeed, the role of gravity in the solvent exchange process in such thick channels could later be confirmed in ref.\ \cite{yu2015}.

The aim of this present paper therefore is to go to a different geometry, namely to a Hele-Shaw cell: a channel formed by two closely spaced parallel glass plates. The employed Hele-Shaw cell possesses a much smaller cell height of $h = 100\mu m$, implying an Archimedes number $Ar = g h^3 \Delta \rho /( \nu^2 \rho )$(see ref.\ \cite{yu2015}) of $Ar =  0.61$, where $g = 9.81 m/s^2$ is the gravitational acceleration, $\nu = 10^{-6} m^2/s$ is the kinematic viscosity, $\rho = 1000 kg/m^3$ is the density of the fluid for injection (deionized water, in this case), and $\Delta \rho = 62 kg/m^3$ is the density difference between a solution to be repelled from the Hele-Shaw cell  (water-ethanol solution with the ethanol concentration of $30$ $vol\%$, in this case) and the fluid for injection. For $Ar < 1$, gravitational effects can be neglected. The Hele-Shaw geometry has the additional advantage that the flow rate now depends on the radial distance $r$ from the point of flow injection, allowing for continuous variation of the local flow velocity, which again leads to the continuous tuning of the Peclet number, which is again the non-dimensionalized flow rate, here given by
\begin{equation}
Pe = {Q \over 2\pi r D}.
\label{pe}
\end{equation}
As a result, the correlation of nanodroplet formation with Peclet number can be easily investigated. The main question which arises is: Does the relationship $\left.\langle Vol\right.\rangle_{area}  \sim$ $h^{3}Pe^{3/4}$ also {\it locally} hold in the Hele-Shaw geometry and is thus universal?

\begin{figure}
	\centering{\includegraphics[width=0.7\textwidth]{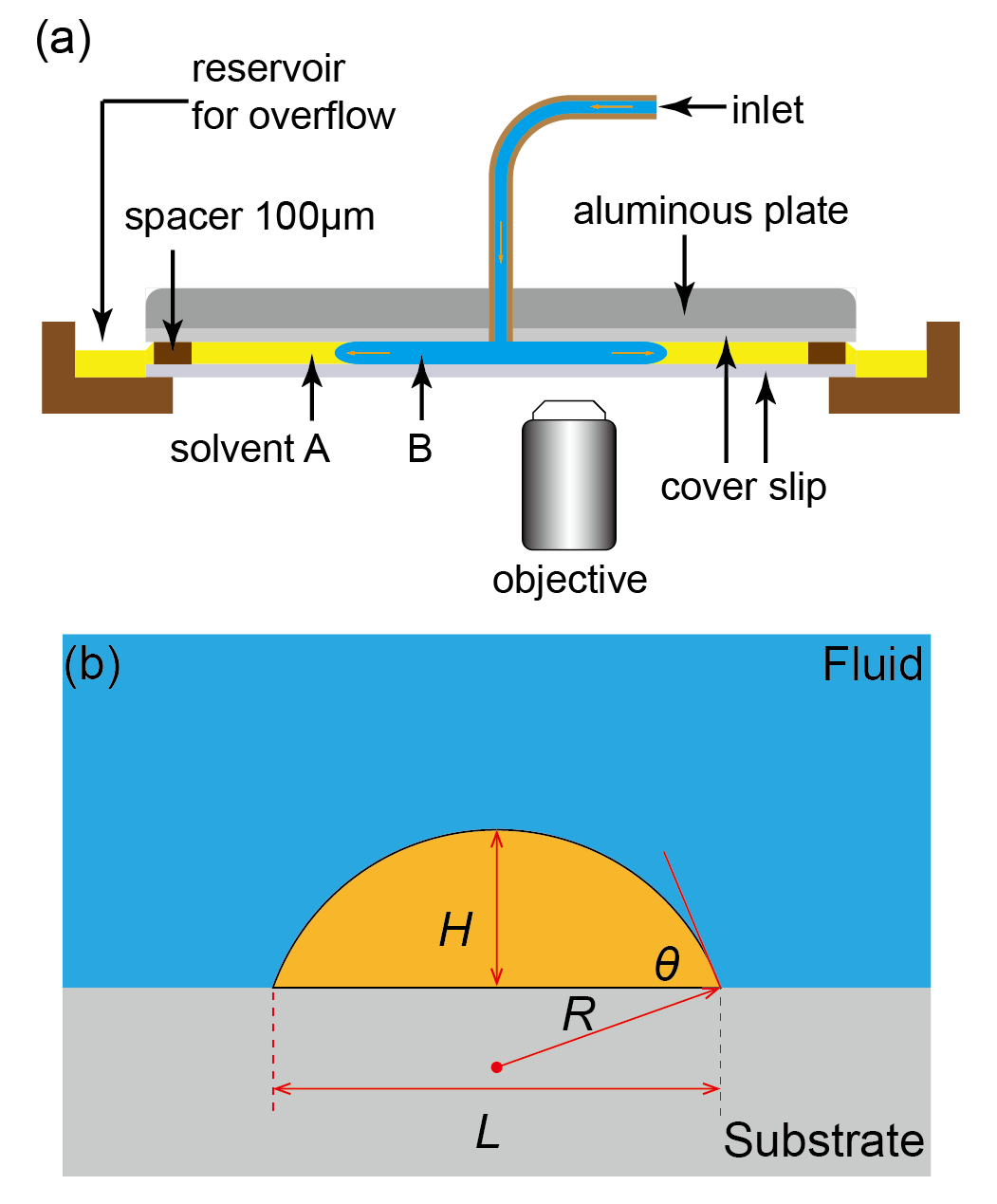}}
\caption{(a) Schematic drawings of the employed Hele-Shaw cell. The cell consists of a bottom cover slip window, a top cover slip attached to an aluminous plate, a spacer of $100  \mu m$ in height, an inlet for liquid and a reservoir for overflow. During experiment, the liquid injected through the inlet will flow outwards in radial direction and is observed from the bottom. (b) Schematic diagram of a nanodroplet, where $R$, $\theta$, $H$, and $L$ are the radius of curvature, the contact angle, the height, and the footprint diameter of the droplet.}
\label{sketch}
\end{figure}

Figure \ref{sketch} shows a sketch of the geometry and the employed notations for the nanodroplets. With this setup, we want to test the fluid dynamical theory of solvent exchange (which can straightforwardly be generalized to the present circular geometry) developed in reference \cite{zhang2015pnas}.

The paper is organized as follows: Section 2 gives the details on the employed method and how the data were collected. Section 3 shows the results, followed by a discussion and the  conclusions (section 4).

\section{Experimental Section} \label{method}
	\subsection{Sample preparation}\label{preparation}
A circular glass cover slip (GOLO, China) with a diameter of 50 mm was used as the substrate for droplet nucleation and bottom window for observation in the Hele-Shaw cell. The substrate was hydrophobilized by PVDF-HFP (poly(vinylidene fluoride-co-hexafluoropropylene), Mw = 400000, Sigma-Aldrich, USA). To do so, the cover slip was first cleaned in a sonication bath of piranha solution (70\% H$_{2}$SO$_{4}$- 30\% H$_{2}$O$_{2}$ solution) for 30 min, followed by the sonication bath of acetone and ethanol for 30 min, then in deionized water 3 times for 5 min. After that, the sample was dried by nitrogen gas. The dried cover slip was then immersed in a 3 $\mu L$ PVDF-HFP in a petri dish. The petri dish remained in an oven at 150$^{\circ}$C for 12h. Eventually, a hydrophobic glass substrate coated with PVDF-HFP was obtained. The measured static contact angle of water on the obtained PVDF-HFP hydrophobic surface is 110$^{\circ}$.
	\subsection{Formation of nanodroplets through solvent exchange in the Hele-Shaw geometry}\label{solvent-exchange}
	Nanodroplets of trans-anethole (4-Propenylanisole, trans-1-Methoxy-4-(1-propenyl)benzene, Solarbio, USA) were produced on the obtained hydrophobic substrate in a Hele-Shaw cell through solvent exchange. As shown in Figure \ref{sketch}a, the top cover slip and the bottom cover slip window form a disk-shaped channel with a height $h = 100\mu m$ and a diameter $d = 50mm$. Unlike the liquid cell applied in ref.\ \cite{zhang2015pnas}, the inlet of the Hele-Shaw cell is in the center of the cell. The size and area density of the droplets are influenced by the solution composition for the solvent exchange \cite{lu2016,Li2018}. In our experiments, two solutions, solvent A and solvent B, were prepared for solvent exchange. Solvent A is an aqueous ethanol solution with the ethanol (analytically pure,99.8$\%$, Aladdin, China) concentration of $30\%$ $vol$. The solution was saturated with trans-anethole, which was labeled by a fluorescence dye perylene (Klamar-reagent, China). Solvent A serves as good solvent for trans-anethole with saturation concentration of $0.55$ $wt\%$. Solvent B is deionized (DI) water, which serves as poor solvent for the trans-anethole. Before the experiments, solvent B was saturated with trans-anethole as well. Such solution composition produced desirable size and density of droplets, suitable for optical images and data analysis. During the solvent exchange process, $200\mu L$ solution A was displaced by 2 mL solution B. The injection of solution B was performed by a syringe pump at three different flow rates of $Q_1 = 1000$, $Q_2 = 500$, and $Q_3 = 300\mu L/min$.
	\subsection{Characterization of nanodroplets}\label{characterization}
	After the tran-anethole droplets had nucleated on the hydrophobic glass substrate, they were characterized by a reflecting fluorescence microscope (IX71, Olympus, Japan). Two objectives ($4\times$ and $20\times$) were used in the imaging of the nucleated droplets. The images taken with the 20$\times$ objective were used for morphological analysis, while the ones taken with the 4$\times$ objective were used to determine the radial distance $r$ for each selected area in the the optical images taken with the 20$\times$ objective. All the optical images ($20\times$) were analyzed using a home designed image segmentation algorithm for the optimized extraction of the droplet footprint diameters. For details of the algorithm, readers are referred to our previous publications \cite{wang2015bjon,Wang2015Segmentation,Wang2018Automated,wang2017}.

For each of the three flow rates, the nucleated nanodroplets with different radial distance $r$ were captured. The number of the captured nanodroplets are 18332, 14403, and 19990 for flow rates of $1000\mu L/min$, $500\mu L/min$, and $300\mu L/min$, respectively. For each $Q$,  the captured droplets were binned in concentric circular bands with a band width $\Delta r = 100\mu m$.

To extract volumes of the nucleated nanodroplets, a tapping mode atomic force microscopy (TM-AFM) (Resolve, Bruker, USA) was also applied to get high-resolution three dimensional (3D) images of the nucleated droplets, after the growth of the oil droplets was already finished. The nucleated nanodroplets remained in solvent B and were imaged in the liquid mode of TM-AFM. For the duration of the operational time, we did not observed any temporal evolution in the droplet lateral size. The obtained AFM images were analyzed with a home-designed Matlab program to extract the height, width and contact angle of the droplets, as reported in our previous work \cite{Wang2017Entrapment}. After that, the volumes of the nanodroplet can be obtained.

Since the number of the captured nanodroplets is huge, it is impractical to measure the volumes of all individual nanodroplets with AFM. Therefore, in this study, only a few nanodroplets with different footprint diameters were imaged with TM-AFM, in order to establish the dependence of the nanodroplet height $H$ and contact angle on its footprint diameter. The height of droplets $H$ was measured for droplets with different footprint diameter $L$, as shown in figure \ref{afm}b. From the $H(L)$ relationship, we obtained the $\theta(L)$ dependence, as shown in the dotted blue curve, which provides the basis for estimation of the contact angle for droplets with different $L$ determined from the optical images. This dependence was further used to obtain the volume for each captured nanodroplet in the optical images from the footprint diameter.

	\subsection{Fluid dynamics theory of solvent exchange}\label{model}
	The scaling law between the final area-averaged volume of the droplet  $\left.\langle Vol\right.\rangle_{area}$ and the Peclet number $Pe$ was established in ref.\ \cite{zhang2015pnas}. When gravitational effect in solvent exchange process can be ignored (i.e., for small Archimedes number), the final area-averaged droplet volume scales as:
\begin{equation}
\left.\langle Vol\right.\rangle_{area} \sim h^3 \left(\frac{C_{sat}} {\rho _ {oil}}\right)^{3/2}\left(\frac{C_{\infty}} {C_{sat}} - 1\right) Pe^{3/4}
\label{vol-pe}
\end{equation}
where $C_{sat}$ and $C_{\infty}$ are the saturation concentrations of oil in the poor solvent and the actual oil concentration, respectively, and $\rho _ {oil}$ is the oil density. This fluid dynamics theory of solvent exchange will be further experimentally validated in this paper.

\begin{figure}[h]
\centering{\includegraphics[width=0.75\textwidth]{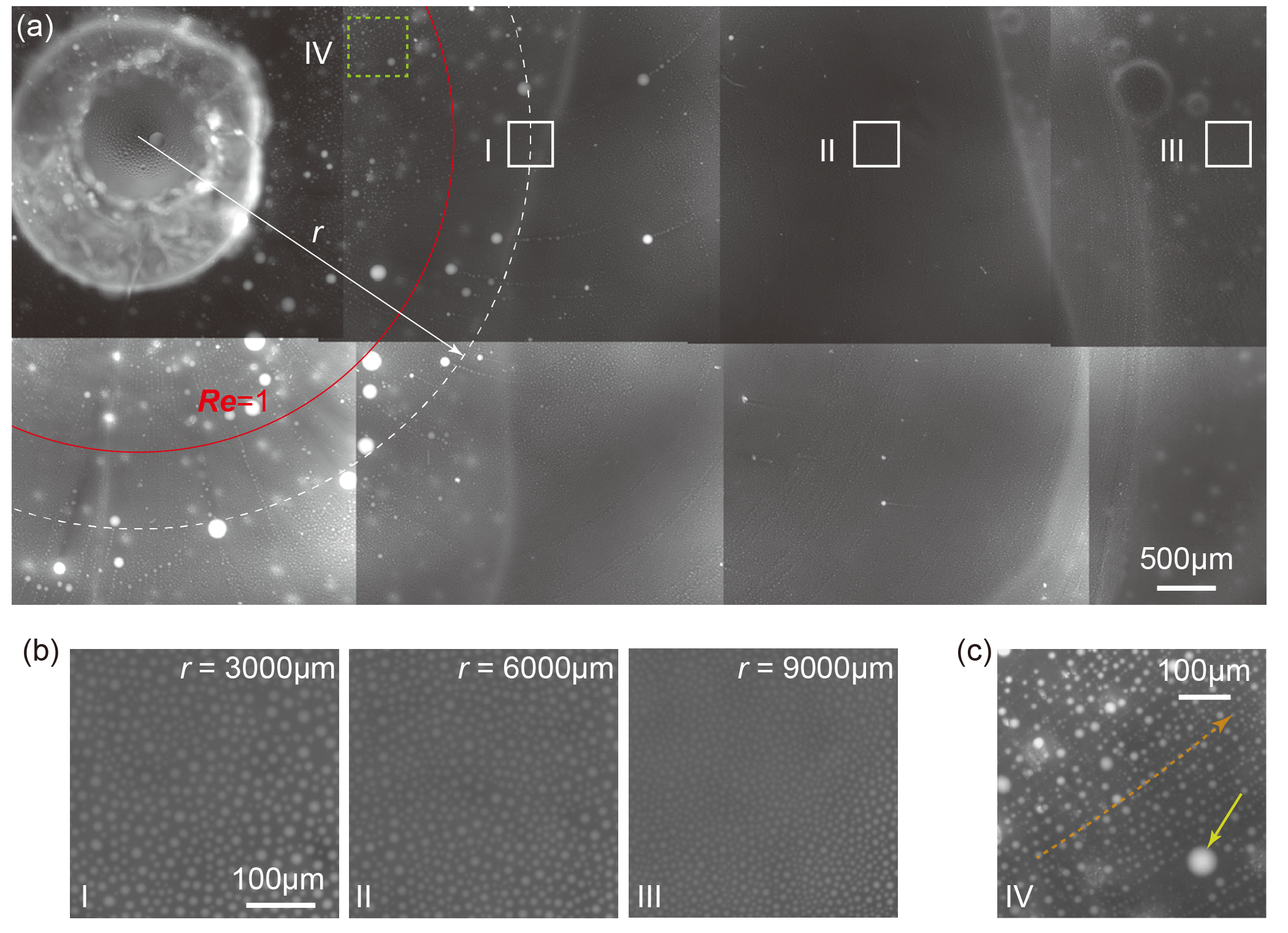}}
\caption{Morphological characterization of the nucleated nanodroplets in the Hele-Shaw geometry: (a) An optical microscope image obtained by aligning different images taken at different radial distance $r$ (4$\times$ objective) for $Q = 1000\mu L/min$. (b) Three example images taken with the 20$\times$ objectives at the three selected positions in white boxes of I, II, and III in (a). (c) An enlarged image for an area selected by a dotted green box IV in (a).}
\label{view}
\end{figure}

\section{Results and Discussion} \label{results}
	After solvent exchange, nucleated nanodroplets were attached to the bottom cover slip window. Figure \ref{view}a shows an fluorescence image obtained by aligning different images taken with the 4$\times$ objective at the flow rate of $Q = 1000\mu L/min$. Three examples of images taken with a higher magnification objective of 20$\times$ are shown in figure \ref{view}b for the three selected areas in figure \ref{view}a. From the images, one can clearly see that the droplet size decreases with increasing distance $r$ from the center. Figure \ref{view}c shows an enlarged image for an area selected by the dotted green box in figure \ref{view}a. One can see that the droplets in the area are lined up along the flow direction (as guided by a dotted orange arrow) and exhibits nonuniform size distributions. This will be discussed later in this section.

\begin{figure}[h]
\centering{\includegraphics[width=0.5\textwidth]{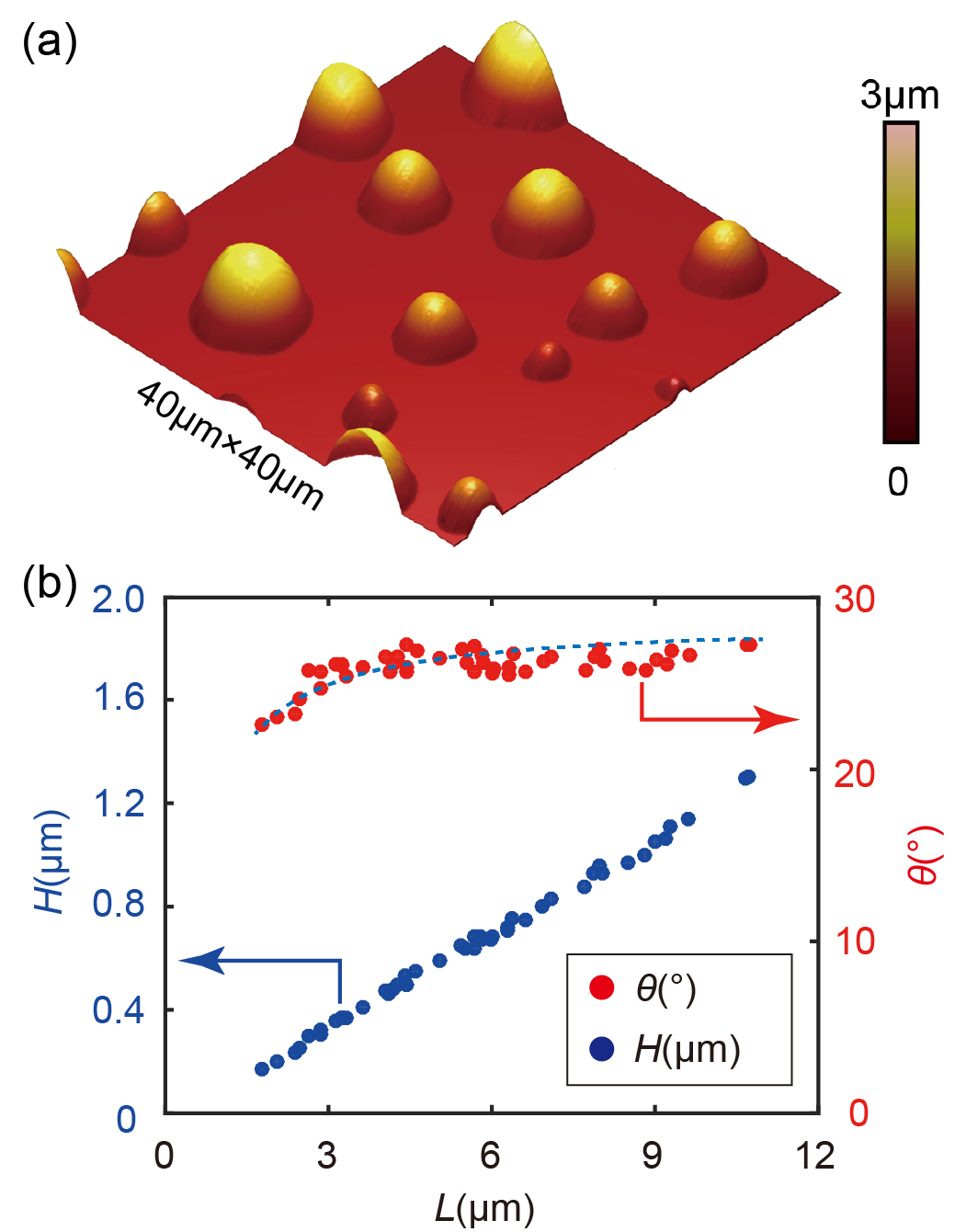}}
\caption{(a) A TM-AFM image of the nucleated droplets, providing their high-resolution 3D morphological characterization. (b) The height $H$ and the contact angle $\theta$ as functions of the footprint diameter $L$ of the droplets. The dependence were further used to obtain $H$ and $\theta$ for nanodroplets in the optical images.}
\label{afm}
\end{figure}

Figure \ref{afm}a shows a representative AFM image of the trans-anethole droplets scanned in solvent B on the bottom substrate. From the AFM images, the height and the contact angle for the nanodroplets were extracted, as shown in figure \ref{afm}b. One can see that the droplet height $H$ roughly linearly increases from about $0.2 \mu m$ to $1.3\mu m$ as $L$ increases from about $1.5\mu m$ to $11\mu m$. The contact angle $\theta$ basically remains constant at $27^{\circ} $ after $L$ is larger than $3\mu m$. The aspect ratios of the nucleated droplets are consistent with the results reported in ref.\cite{zhang2015,zhang2015pnas} for the same combination of solvents and solutes.

\begin{figure}[h]
	\centering{\includegraphics[width=0.7\textwidth]{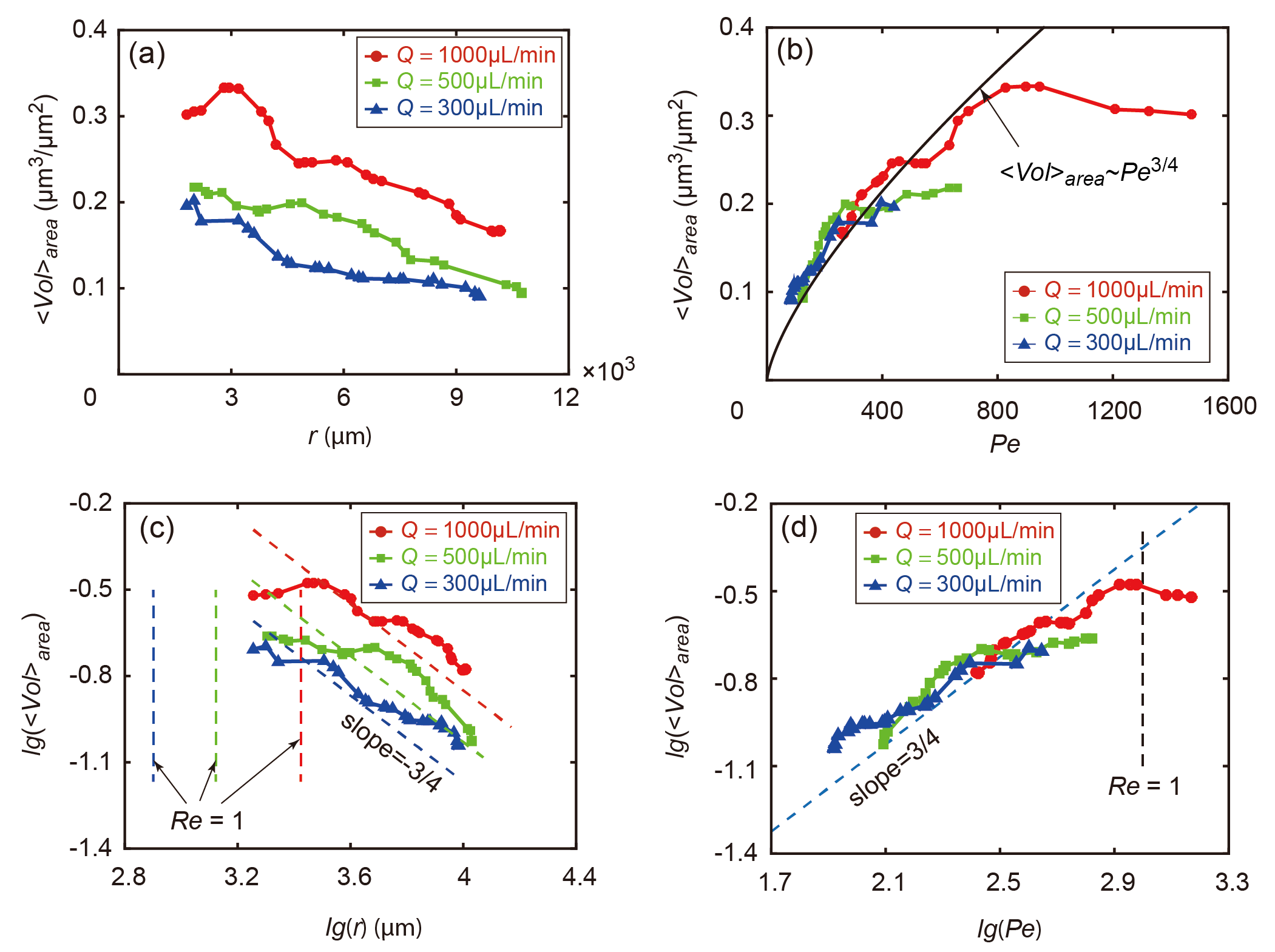}}
\caption{(a, c) Linear and double-logarithmic scale plots of the mean volume per area $\left.\langle Vol\right.\rangle_{area}$ of the nucleated nanodroplets as a function of radial distance $r$ from the flow inlet for the three different flow rates $Q$ of $1000\mu L/min$, $500\mu L/min$, $300\mu L/min$ (top to bottom). (b, d) Linear and double-logarithmic scale plots of $\left.\langle Vol\right.\rangle_{area}$ as a function of the dimensionless form of flow velocity -- Peclet number $Pe$ for different $Q$. All three curves collapse on a curve with a slope of 3/4 (shown as straight line) in the log-log plot, indicating an universal scaling law of $\left.\langle Vol\right.\rangle_{area} \sim Pe^{3/4}$.}
\label{mean-vol}
\end{figure}

	With the developed dependence between $H$, $\theta$, and $L$, the droplet volumes $Vol$ for individual captured droplets in the optical microscope images were obtained. 
After that, the mean droplet volume per area $\left.\langle Vol\right.\rangle $ $_{area}$ was further calculated, as shown in figure \ref{mean-vol}a. For all the three flow rates, $\left.\langle Vol\right.\rangle _{area}$ decreases with increasing $r$. Moreover, for the same $r$, $\left.\langle Vol\right.\rangle _{area}$ increases with increasing flow rates. Most importantly, regardless of the flow rates, all the three curves are well superposed on each other in the plot of $\left.\langle Vol\right.\rangle_{area} $ versus Peclet number $Pe$, as shown in figure \ref{mean-vol}b, especially for lower $Pe$ value (corresponding to larger distance $r$ from the inlet).

The double logarithmic plots of $\left.\langle Vol\right.\rangle_{area}$ versus the radial distance $r$ and {\it vs} the Peclet number $Pe$ are shown in figure \ref{mean-vol}c,d. It is clear that $\left.\langle Vol\right.\rangle_{area}$ shows a -3/4 and 3/4 power law dependence on $r$ and $Pe$, respectively. Most importantly, three curves in figure \ref{mean-vol}d collapse on one universal curve. These results are consistent with the theoretical prediction (eq.\ (\ref{vol-pe})) and provide strong support to the model.

In figure \ref{mean-vol}c, d, for smaller $r$ or larger $Pe$, the data points for $Q=1000 \mu L/min$  start to deviate from the -3/4 or 3/4 power law scaling lines. The deviation is likely due to the higher Reynolds number corresponding to smaller $r$. In the Hele-Shaw cell, the Reynolds number $Re$ at the radial distance $r$ is given by
\begin{equation}
Re = \frac{Q}{2\pi rv}.
\label{re}
\end{equation}
where $v$ is the kinematic viscosity of the solution (here $v=10^{-6} m^2/s$ for water). In  figure \ref{mean-vol}c, the red, green, and blue vertical dashed lines correspond to $Re = 1$ for the flow rates of $1000\mu L/min$, $500\mu L/min$, and $300\mu L/min$, respectively. In figure \ref{mean-vol}d, a dashed vertical line was also drown at the position of $Re = 1$. One can see that the deviations mainly occur in the area where the Reynolds number $Re > 1$. Indeed, at higher Reynolds numbers, the flow becomes dominated by inertia effects and the laminar theory of ref\cite{zhang2015pnas} would require extensions. Namely, vortical flow structures may occur, leading to a less organized flow pattern. This may lead to advection of nucleated droplets.

For that case $Re>1$, droplets indeed can move outwards along the radial direction of the Hele-Shaw cell. This can be seen from figure \ref{view}c. It is an enlarged image for the area highlighted by the dotted green box in the region corresponding to $Re>1$, which is defined by a red circle in figure \ref{view}a. In the image, one can see that the droplets are mostly lined up along the radial direction, which is clearly different from that shown in figure \ref{view}b. The advected droplets can merge with other surface droplets on their paths, leading to an increased size of droplets, as pointed at by the arrow in figure \ref{view}c. Meanwhile, the strong outward motion of droplets unavoidably leads to a reduced average volume $\left.\langle Vol\right.\rangle _{area}$ close to the center flow inlet. As a result, in the region with $Re>1$, $\left.\langle Vol\right.\rangle _{area}$ is smaller than what is predicted by the model equation(\ref{vol-pe}) of reference\cite{zhang2015pnas} and the experimental results deviate from the power law line, as shown in figure \ref{mean-vol}c,d.

 \begin{figure}[h]
	\centering{\includegraphics[width=0.65\textwidth]{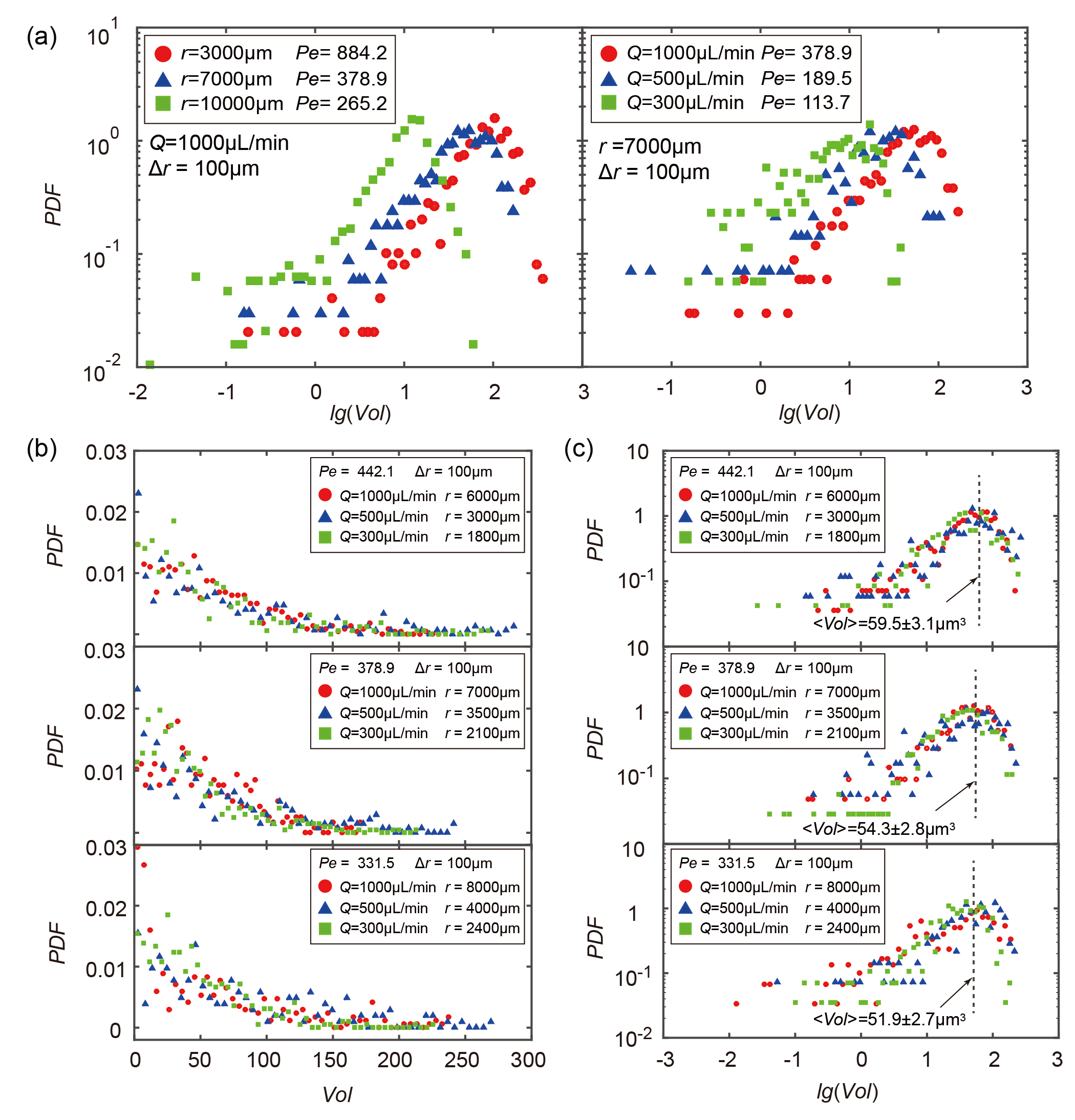}}
\caption{  Droplet volume PDFs for various Peclet numbers $Pe$. (a) Left: PDF of the droplet volume for three different distances $r$ from the point of flow injection at the same flow rate $Q$ of $1000\mu L/min$. Right: PDFs of the droplet volume at the same radial distance $r$ for three different flow rate $Q$. For both cases, the PDFs shift rightwards with increasing $Pe$. (b, c) Comparison of the PDFs of droplet volume in linear (b) and double logarithmic (c) plots for droplets nucleated in the areas with the same local $Pe$ but different combinations of $Q$ and $r$. One can conclude that the PDFs of the droplet volume strongly depend on the local $Pe$, regardless of $Q$ and $r$.}
\label{pdfs}
\end{figure}

Compared to the result reported in ref.\ \cite{zhang2015pnas} (Figure 2G, therein), the result shown in figure \ref{vol-pe}d provide a higher consistency among the experiments with different flow rates. This is attributed to the smaller cell height of $h = 100\mu m$, which eliminates gravitational effects (i.e., Ar < 1). The results thus give strong support to our previously developed fluid dynamics model of solvent exchange, namely, that the volume of the nucleated droplets strongly depends on the dimensionless flow velocity, perfectly following the relationship $\left.\langle Vol\right.\rangle_{area}  \sim$ $h^{3}Pe^{3/4}$.


After having shown the universality of the 3/4 power law dependence between $\left.\langle Vol\right.\rangle_{area}$ and $Pe$, one also wonders on the universality of the probability distribution function (PDF) of the droplet volumes, i.e., on the dependence of the PDFs on $Pe$. To answer this question, the distribution of the droplet volume was calculated. Figure \ref{pdfs}a shows the PDF of the droplet volume at different $r$ for the flow rate $Q = 1000\mu L/min$ (left figure) and at the same radial distance $r = 7000\mu m$ for different flow rates (right figure). In the left figure, the PDF of droplet volumes shifts rightwards with decreasing $r$. Similarly, the PDF shifts rightwards with increasing $Q$ in the right figure. Both figures of course reflect that the mean droplet volume increases with $Pe$.

Since $Pe$ is continuously varying in the Hele-Shaw cell, this gives us the freedom to select a particular $Pe$ value for fixed flow rate. This allows us to compare the distribution of the droplet volume with the same $Pe$ value, but different $Q$. For the three different flow rates, three different local $Pe$ values of $Pe = 442.1$, $378.9$, and $331.5$ were selected. This corresponds to nine areas in total, three for each of flow rate. The linear and double-logarithmic plots of the PDF of the droplet volumes for the three selected $Pe$ values are shown in figure \ref{pdfs}b and c, respectively. Remarkably, it shows that the PDFs of the droplet volume have an {\it universal} dependence on the Peclet number, regardless of the radial distances $r$ and flow rates $Q$.

\section{Conclusions} \label{conclusions}
	In summary, we experimentally investigated the formation of surface nanodroplets by solvent exchange under a well controlled flow conditions. Compared to the rectangular cross section channels used in one of our previous study \cite{zhang2015pnas}, a Hele-Shaw cell with a cell height of 100 $\mu m$ was employed. In the new setup, gravitational effects can be negligible. Moreover, the Hele Shaw setup easily allows the continuous tuning of the dimensionless flow velocity - namely the Peclet number $Pe$. By combining a fluorescence optical microscope and an AFM, the height, contact angle, footprint diameter, and volume of surface nanodroplets were extracted for a huge amount of nanodroplets under three different flow rates. The results reveal the underlying mechanism governing the droplet nucleation through solvent exchange. They show that not only the mean droplet volume, but also the PDF of droplet volumes universally depends on the local $Pe$ number. Although the size of the nucleated droplets changes with radial distance $r$ and flow rate $Q$, the mean droplet volume per area $\left.\langle Vol\right.\rangle_{area}$ shows a universal 3/4 power law dependence on $Pe$. This is in a good agreement with the model developed in our previous work \cite{zhang2015pnas}. Moreover, further investigation shows that the PDFs of the droplet volume also follow an universal dependence on the local $Pe$, regardless of the radial distance $r$ and the employed flow rate $Q$. The revealed dependences provide an important guideline for the control of the flow conditions in the mass-production of surface nanodroplets, which is very relevant for various applications, such as in diagnostics, liquid-liquid microextraction, drug production, or food-processing.

\begin{acknowledgement}

This work is supported by National Natural Science Foundation of China (Grant No. 51775028), Beijing Natural Science Foundation (Grant No. 3182022), and ERC-NSFC joint program (Grant No. 11811530633). The authors thank the Dutch Organization for Research (NWO) and the Netherlands Center for Multiscale Catalytic Energy Conversion (MCEC) for financial support. D.L. also acknowledges financial support by an ERC-Advanced Grant and by NWO-CW.

\end{acknowledgement}





\bibliography{achemso-demo}

\end{document}